\documentclass[preprint,showpacs]{revtex4}
\usepackage{epsfig}
\usepackage{graphicx}
\usepackage{bm}
\usepackage{amssymb}
\newcommand{\bmath}[1]{\mbox{\boldmath{${#1}$}}}

\newcommand{\half}{\mbox{${\textstyle \frac{1}{2}}$}}           
\newcommand{\eightth}{\mbox{${\textstyle \frac{1}{8}}$}}         

\newcommand{\rd}{\textrm{d}}
\begin{document}
\title{Meson production in high-energy electron-nucleus scattering}
\date{\today}
\author{G\"oran F\"aldt}\email{goran.faldt@physics.uu.se}  
\affiliation{ Department of physics and astronomy, \
Uppsala University,
 Box 516, S-751 20 Uppsala,Sweden }

\begin{abstract}
Pseudoscalar mesons can be produced and studied in  high-energy 
electron-nucleus scattering. We review and extend our previous  analysis of 
meson production in the nuclear Coulomb field. 
The $P\rightarrow \gamma \gamma$ decay rates  
are most directly determined for mesons produced in the double-Coulomb region where both
photons are nearly real, and provided
the background-hadronic contribution remains small. 
The larger
the mass of the meson the higher  the electron energy needed to
assure such a condition. 

\end{abstract}
\pacs{24.10.Ht, 25.20.Lj, 25.30.Rw}
\maketitle
%
%
%
\section{Introduction}\label{ett}

Pseudoscalar mesons are produced in high-energy electron-nucleus 
scattering through a two-photon process where the two photons are 
radiated, one by the high-energy electron and the other by the atomic nucleus.
This possibility was studied, in the  Born approximation,
 by Hadjimichael and Fallieros \cite{Hadji}. Recently,
  a Glauber-model \cite{RJG} description  of the 
  same process has been developed \cite{GFI}.
 Theoretical studies of the two-photon-fusion process are 
 particularly valuable   
since the PrimEx Collaboration \cite{PrimEx} aims at measuring electromagnetic 
properties of pseudoscalar mesons through  this effect, 
in 11 GeV/$c$ electron-nucleus scattering.

The kinematics of the electron-nucleus-meson-production
 reaction is defined through
\begin{equation}
e^-(k_1)+ {\rm A}(p_1)\rightarrow e^-(k_2)+P(K)+{\rm} A(p_2),
\end{equation}
where $P$ represents one of the mesons $\pi^0$, $\eta$, or $\eta'$.
Our analysis is relevant for high electron and meson energies and small transverse
momenta.
In addition, the momentum transfers to meson and nucleus should
preferentially be in the double-Coulomb region, leading to further  
restrictions. The mechanism dominating  this region is, at sufficiently high energies, 
bescribed by the graph of  Fig.1. 
\begin{figure}[ht]
\scalebox{0.90}{\includegraphics{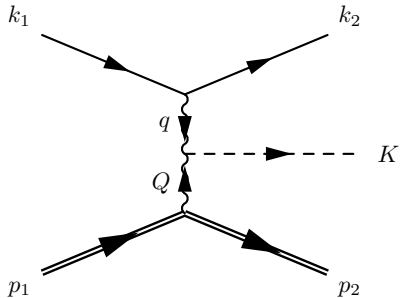} }
\caption{Graph describing pseudoscalar-meson production in the Coulomb field of 
a nucleus in electron coherent-nucleus scattering.}\label{F1-fig}
\end{figure}

This graph shows that pion-nucleus photoproduction is a
subgraph of the pion-nucleus electroproduction, but only in special kinematic circumstancies
does the photoproduction amplitude enter as a separate factor. Out treatment
of the electroproduction amplitude is similar to our previous treatment 
of the photoproduction amplitude \cite{GF-bas}.

The  cross-section distribution 
is in the Coulomb region mainly determined by the photon propagators,
which lead to structures like
\begin{equation}
	\frac{{k}_{\bot}}{\mathbf{k}_{\bot}^2 + k_{\|}^2/\gamma^2 },
\end{equation}
with $\mathbf{k}_{\bot}$ the variable  photon transverse momentum,
$k_{\|}$ the fixed  photon longitudinal momentum, and $\gamma$ the gamma factor
of the radiating charge. 
This behaviour results in a cross-section maximum at ${k}_{\bot}=k_{\|}/\gamma$, 
the Primakoff peak.
For the low-energy photon radiated by the nucleus $k_\|/\gamma=m_P^2/2K_\|$,
where $m_P$ is the mass of the meson and $K_\|=\mathbf{K}\cdot\hat{k}_1$ its longitudinal 
(or total) momentum. This 
expression  is well known from ordinary Coulomb production. 
For the high-energy photon radiated by the high energy electron the 
effective longitudinal momentum  is $k_\|/\gamma=m_e$, with 
$m_e$ the electron mass.

In the PrimEx experiment \cite{PrimEx} typical energies are;
for the incident electron $E(\mathbf{k}_1)=11$ GeV, 
for the scattered electron $E(\mathbf{k}_2)=300$ MeV, and  
for the pseudoscalar meson
$\omega(\mathbf{K})=10.7$  GeV.  Consequently,  the energy of
the virtual photon radiated by the electron, and initiating the meson production through fusion
with a soft photon radiated by the nucleus, is also 10.7 GeV.
These numbers are only for numerical illustration. The model we present is in
itself a general one.
 
The electron-coherent-nucleus-production amplitude is a sum of two amplitudes;
the two-photon-fusion amplitude ${\cal M}_{2\gamma}$ of Fig.1, and the
electron induced hadronic-photoproduction amplitude $\cal{M}_{\gamma}$ of Fig.2.
\begin{figure}[ht]
\scalebox{0.90}{\includegraphics{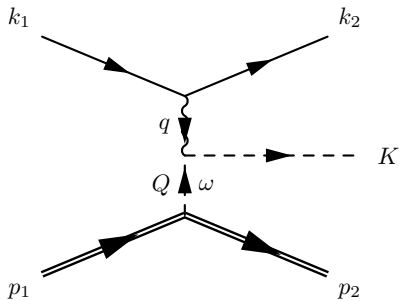} }
\caption{Graph describing the hadronic contribution to pseudoscalar-meson product
in electron coherent-nucleus scattering. Dominant contribution from $\omega$ exchange.}\label{F2-fig}
\end{figure}

Normalizations are chosen such that in the Born approximation 
the unpolarized 
 cross-section distribution for the two-photon-fusion contribution  simplifies to
\begin{equation}
\frac{\rd \sigma}{\rd^2k_{2\bot}  \rd^2K_{\bot} \rd k_{2\|}}
  = \frac{1}{\pi K_\|  }
    \bigg[ \frac{Z\alpha^2}{m_\pi}g_{\pi\gamma\gamma}\ \frac{q_{\bot}}{\mathbf{q}_\bot^2+m_e^2}
    \cdot\frac{Q_{\bot}}{\mathbf{Q}_\bot^2+Q_\|^2}\bigg]^2  ,
    \label{Born-cross}
\end{equation}
with $-Q_\|=m_P^2/2K_\|$.
The  structure of the cross-section distribution at 
small-transverse momenta is essentially determined by
the photon-exchange propagators. There are two such propagators;
one in the variable $\bmath{q}_{\bot}$ and one in the variable $\bmath{Q}_{\bot}$.
Each of them exhibits a Primakoff-peak structure. We observe that 
the spin-averaged Born approximation does not depend on
the angle between the transverse momenta $\mathbf{q}_\bot$ and $\mathbf{Q}_\bot$.

The numerical illustrations in Ref.\cite{GFI} erroneously employ positive values
for $Q_\|$. However, this error is of no concern since the quantities graphed are 
independent of the sign of $Q_\|$.

%
%
%
%
\newpage
\section{Classical radiation by a relativistic electron}

The pseudoscalar-meson decay $P\rightarrow \gamma\gamma$   is governed by the
Hamiltonian density
\begin{equation}
	{\cal H}(x)=\eightth g\,  F(x)\cdot\tilde{F}(x)P(x), \label{decay-H}
\end{equation}
where ${F}_{\mu\nu}(x)$ is the electromagnetic-field tensor, $\tilde{F}_{\mu\nu}(x)=\epsilon_{\mu\nu\sigma\tau}F^{\sigma\tau}$
its dual,
with $\epsilon_{0123}=1$, and $P(x)$ the pseudoscalar-meson field.
The coupling constant $g=e^2 g_{\pi\gamma\gamma}/m_\pi$.
 In terms of the electric and magnetic 
fields the Hamiltonian reads
\begin{equation}
{\cal H}(x)=\half g \, 
 \mathbf{E}(x)\cdot\mathbf{B}(x)P(x).\label{decay-H2}
\end{equation}
In our application the electromagnetic fields are generated by the electron
and nuclear charges. 
It is therefore instructive to study first the classical counterparts 
of these fields \cite{LL}.

A point particle of charge $e$ moves along the trajectory 
$\mathbf{r}(t)=\mathbf{r}_0+\mathbf{v}t$. The associated four-vector potential
$A_\mu (x)$ satisfies the Maxwell equation $\Box A (x)= j(x)$,
with the four-vector current 
\begin{equation}
	j(x)=e(1, \mathbf{v}) \delta(\mathbf{r}-\mathbf{r}_0-\mathbf{v}t).
\end{equation}
In a plane-wave decomposition of the four-vector
potential,
\begin{equation}
	A(x)=\int \frac{\rd^3 k}{(2\pi)^3}                 
	         A(\mathbf{k},t)e^{i\mathbf{k}\cdot\mathbf{r}},
	         \label{A-pot}
\end{equation}
the  solution to the Maxwell equations takes the form
\begin{equation}
	A(\mathbf{k},t)=e(1, \mathbf{v})
	\frac{e^{-i\mathbf{k}\cdot(\mathbf{r}_0+\mathbf{v}t)}}
	   {\mathbf{k}^2-(\mathbf{k}\cdot\mathbf{v})^2}.
\end{equation}
This four-vector potential gives the following 
plane-wave decompositions of the electric and magnetic fields
\begin{eqnarray}
	\mathbf{E}(\mathbf{k},t)&=&ie 
	     \frac{-\mathbf{k}+(\mathbf{k}\cdot\mathbf{v})\mathbf{v}}{\mathbf{k}^2-(\mathbf{k}\cdot\mathbf{v})^2}  
	  e^{-i\mathbf{k}\cdot(\mathbf{r}_0+\mathbf{v}t)\mathbf{v}t}, \label{E-field} \\ && \nonumber \\
	\mathbf{B}(\mathbf{k},t)&=&ie \frac{\mathbf{k}\times\mathbf{v}}   {\mathbf{k}^2-(\mathbf{k}\cdot\mathbf{v})^2}   
	   e^{-i\mathbf{k}\cdot(\mathbf{r}_0+\mathbf{v}t)}.\label{B-field}
\end{eqnarray}
The denominators can be rewritten as 
\begin{equation}
	\mathbf{k}^2-(\mathbf{k} \cdot\mathbf{v})^2=\mathbf{k}_{\bot}^2+
	   k_{\|}^2/\gamma^2, \label{Gamma-den}
\end{equation}
with $\gamma=E/m$  the  relativistic gamma factor of the charged particle.
Transverse and parallel refer to directions orthogonal and parallel
to the velocity $\mathbf{v}$ of the charged particle. The numerators, on the
other hand, can be rewritten as 
\begin{eqnarray}
	\mathbf{k}- (\mathbf{k}\cdot\mathbf{v})\mathbf{v}&=&(\mathbf{k}_\bot,k_\|/\gamma^2)
	 , \label{E-num} \\ 
	\mathbf{k}\times\mathbf{v}&=&\mathbf{k}_\bot\times\mathbf{v}.\label{B-num}
\end{eqnarray}

In the application to the PrimEx experiment two limits are of interest.
The non-relativistic limit, $v\ll 1$, applies to the soft radiation by the nucleus.
This corresponds to cyclotron radiation. The denominators of Eqs (\ref{E-field}) and 
(\ref{B-field}) reduce to $\mathbf{k}^2$.  The other
limit applies to hard radiation by the ultra-relativistic electron,
where, in the denominator,  the longitudinal momentum $k_\|$ is cut down by the Lorentz factor $1/\gamma$.
This corresponds to synchrotron radiation.

In  the PrimEx experiment there are kinematic restrictions.
The longitudinal momentum 
of the soft photon is fixed to be $m_P^2/2K_\|$, with
$K_\|$ the pion momentum, and the longitudinal momentum of the hard photon is very
nearly equal to the momentum of the electron, so that $k_{\|}/\gamma\approx m_e$.
Thus, these two  parameters act as cut-offs for the 
corresponding transverse momentum distributions.  The longitudinal component of
the electric field is for the hard photon cut down by the factor $1/\gamma^2$ and may be neglected.

According to Eq.(\ref{decay-H2}) pion decay is determined by the scalar product 
$\mathbf{E}(x)\cdot\mathbf{B}(x)$. The electric field must be associated with the 
non-relativistic nucleus, 
since its magnetic field is very weak, and the magnetic field consequently 
associated with the ultra-relativistic electron.

%
%
%
\newpage
\section{The Coulomb contribution}

We shall now review the formulae for the Coulomb contribution to electron-induced pion
production, i.e.~the two-photon-fusion amplitude of Fig.~1. The corresponding cross-section
distribution has a double-peak-Primakoff structure. We normalize
so that the cross-section distribution takes  the form
\begin{equation}
\frac{\rd \sigma}{\rd^2k_{2\bot}  \rd^2K_{\bot} \rd k_{2\|}}
  = \frac{1}{\pi K_\|  }
    \bigg[ u^{\dagger}(k_2){\cal M}_{2\gamma} u(k_1)\bigg]^2  ,
\end{equation}
where $u(k_1)$ and $u(k_2)$ are two-component spinors for  
incident and scattered electrons. The two-photon amplitude is
decomposed as 
\begin{eqnarray}
	{\cal M}_{2\gamma}&=&i {\cal N}_{2\gamma}\,
	 \bigg[ 	G(\mathbf{q},\mathbf{Q}) 
	   -   H(\mathbf{q},\mathbf{Q})\ 
	   i\bm{\sigma}\cdot \hat{k}_1 \bigg],
	   	 \label{M2g-decomp} \\
	{\cal N}_{2\gamma}&=& Z\, \frac{\alpha^2 g_{\pi\gamma\gamma}}{m_\pi}  .
	\label{N2g-factor}
\end{eqnarray}
The amplitude $G(\mathbf{q},\mathbf{Q})$ originates with the current part of 
the electron-four-vector current, and $H(\mathbf{q},\mathbf{Q})$ with the spin 
part of the same current. 

Taking into account multiple scattering of both electron and  meson the expressions
for the amplitudes become 
\begin{eqnarray}
	G(\mathbf{q},\mathbf{Q})&=&
	  \int\rd^3r_e \int\rd^3r_\pi\,  e^{i\mathbf{q}\cdot(\mathbf{r}_e -\mathbf{r}_\pi)}
	    e^{-i\mathbf{Q}\cdot\mathbf{r}_\pi}  
	    \exp[i\chi_C({b}_e)-\half \sigma_\pi' T(\mathbf{b_\pi},z_\pi)]\nonumber \\	 &&   
	   \times  \left[\mathbf{E}_e(\mathbf{r}_e -\mathbf{r}_\pi)
	      \times \mathbf{E}_A(\mathbf{r}_\pi)\right]\cdot \hat{k}_1  , 
	      \label{G-general}\\
	   H(\mathbf{q},\mathbf{Q})&=&
	  \int\rd^3r_e \int\rd^3r_\pi\,  e^{i\mathbf{q}\cdot(\mathbf{r}_e -\mathbf{r}_\pi)}
	    e^{-i\mathbf{Q}\cdot\mathbf{r}_\pi}  
	    \exp[i\chi_C({b}_e)-\half \sigma_\pi' T(\mathbf{b_\pi},z_\pi)]\nonumber \\	 &&   
	     \times [\mathbf{E}_e(\mathbf{r}_e -\mathbf{r}_\pi)
	      \cdot \mathbf{E}_A(\mathbf{r}_\pi)] ,\label{H-general}   
\end{eqnarray} 
where $\mathbf{r}_e$ and $\mathbf{r}_\pi$ are the electron and pion coordinates.
The nuclear $\mathbf{E}_A(\mathbf{r})$ and electron $\mathbf{E}_e(\mathbf{r})$ electric
fields are the transverse parts of the corresponding fields. The distortion of the electron-wave function
is described by the Coulomb-phase function $\chi_C({b}_e)$; the distortion of the 
pion-wave function by the nuclear-thickness function $T(\mathbf{b_\pi},z_\pi)$.

In the $G(\mathbf{q},\mathbf{Q})$ amplitude of Eq.(\ref{G-general}) 
we recognize $ \mathbf{E}_A(\mathbf{r}_\pi)$ 
as the nucleus-electric field,  and $\mathbf{E}_e(\mathbf{r}_e -\mathbf{r}_\pi)
\times  \hat{k}_1$ as the electron-magnetic field, all in accordance with
Eq.(\ref{decay-H2}). The $H(\mathbf{q},\mathbf{Q})$ amplitude of Eq.(\ref{H-general}) 
is associated 
with the spin current, and its parity is therefore opposite to that of
the  $G(\mathbf{q},\mathbf{Q})$ amplitude. Hence, it is built 
on the scalar product of the two electric fields.

Next, we give the definitions of the various functions entering the integrands
of Eqs (\ref{G-general}) and (\ref{H-general}).
The transverse-electric field of the electron is
\begin{equation}
	\mathbf{E}_e(\mathbf{r})= \frac{1}{4\pi i}
	\frac{\gamma\mathbf{b}}{[\mathbf{b}^2+\gamma^2z^2]^{3/2}}.
	\label{E-field-electr}
\end{equation}
The $\gamma$ dependence in this expression, which is in coordinate space, is compatible
with the $\gamma$ dependence in momentum space, Eq.(\ref{Gamma-den}). I apologize 
for the unconventional multiplicative constant. The transverse-electric field 
of the nucleus
is in general associated with an extended-charge distribution
$\hat{\rho}_{ch}(r)$, which we normalize to unity,
\begin{eqnarray}
		\mathbf{E}_A(\mathbf{r})&=& \frac{-Q(r)}{4\pi i}
	\frac{\mathbf{b}}{[\mathbf{b}^2+z^2]^{3/2}},
	\label{E-field-nucl}	 \\
	 Q(r)&=&4\pi \int_0^r \rd r' r'^2\hat{\rho}_{ch}(r'),
\end{eqnarray}
and $Q(\infty)=1$. For a uniform-charge distribution with radius $R_u$
\begin{equation}
	Q(r)=
	 \left\{
	\begin{array}{l}
	 1,\ r>R_u, \\ 
	  (r/R_u)^3, \ r<R_u.
	   \end{array}\right.
	\label{Qch-ext}
\end{equation} 
Note that the fields of Eqs (\ref{E-field-electr}) and (\ref{E-field-nucl}) are defined 
to have opposite signs.

The expression for the  Coulomb-phase function $\chi_C(b)$ of
 in Eqs (\ref{G-general}) and (\ref{H-general}) has been given by Glauber  
 \cite{RJG, GF-Coul}. We employ the 
integral representation
\begin{equation}
	\chi_C(b)=-2\frac{Z \alpha}{v}  
	\int \rd^3r'\hat{\rho}_{ch}(r')\left[ \ln\left(\frac{\mathbf{b}-\mathbf{b}'}{2a}\right)\right],
	 \label{log_Coul}
\end{equation}
where $a$ is a cut-off parameter common to all amplitudes.
For a point-charge distribution we conclude that
\begin{equation}
	\chi_{pc}(b)=-2\frac{Z\alpha}{v} \ln(b/2a) .
\end{equation}
Additional information on the Coulomb-phase function can be found in Ref.\cite{GF-Coul}, 
in particular expressions for the phase function for uniform-charge distributions.

The pion distortion in Eqs (\ref{G-general}) and (\ref{H-general}) is
controlled by the parameter 
$\sigma_\pi'=\sigma_\pi(1-i\alpha_\pi)$ with $\sigma_\pi$ the pion-nucleon-total-cross section. 
The target-thickness function $T(\mathbf{b},z)$ is defined as
the integral along the pion trajectory of the nuclear-hadronic-matter density 
$A\hat{\rho}_{hd}(r)$ as seen by the pion,
\begin{equation}
	T(\mathbf{b},z)= A \int_z^\infty \rd z'\hat{\rho}_{hd}(\mathbf{b},z'),\label{T-def}
\end{equation}
for a pion produced at $(\mathbf{b},z)$. The density $\hat{\rho}_{hd}(\mathbf{r})$ is normalized to
unity.

The amplitudes $G(\mathbf{q},\mathbf{Q}) $ and $H(\mathbf{q},\mathbf{Q}) $
can be written on a more convenient form by making use of 
the operator replacements, $\bm{b}_e \rightarrow -i\bm{\nabla}_q$ and 
$\bm{b}_\pi\rightarrow +i\bm{\nabla}_Q$, as described in Ref.\cite{GFI}. This
operation leads to the decomposition
\begin{eqnarray}
G(\mathbf{q},\mathbf{Q}) &=& K(\mathbf{q},\mathbf{Q})\ 
	 (\hat{\mathbf{q}}_\bot\times\hat{\mathbf{Q}}_\bot)\cdot\hat{k}_1  
	 + L(\mathbf{q},\mathbf{Q})\ \hat{\mathbf{q}}_\bot\cdot\hat{ \mathbf{Q}}_\bot 
	       ,\label{Gampdecomp}	\\
H(\mathbf{q},\mathbf{Q} )&=& K(\mathbf{q},\mathbf{Q})\ 
      \hat{\mathbf{q}}_\bot\cdot\hat{ \mathbf{Q}}_\bot - 
	  L(\mathbf{q},\mathbf{Q}) \ (\hat{\mathbf{q}}_\bot\times\hat{\mathbf{Q}}_\bot)\cdot\hat{k}_1 
	    \label{Hampdecomp}. 	
\end{eqnarray}

The expressions for the scalar amplitudes $K(\mathbf{q},\mathbf{Q})$ and $L(\mathbf{q},\mathbf{Q})$ are, with $q_\|/\gamma\approx m_e$,
\begin{eqnarray}
	K(\mathbf{q},\mathbf{Q})&=&  
	  \frac{m_eQ_\|}{(2\pi)^2} 
	  \int_0^\infty b_e\rd b_e \ K_1(m_eb_e) \int_0^\infty b_\pi \rd b_\pi \ I(b_\pi,  Q_\|)
	  \nonumber \\ &&
	 \times  2\pi \int_0^{2\pi} \rd \phi 
	  \, J_0\left( \sqrt{X} \right)
	\{\cos\phi\} \exp[i\chi_C({B}_e] , \label{Kampffint} \\ && \nonumber \\
 	L(\mathbf{q},\mathbf{Q})&=&  
	  \frac{m_eQ_\|}{(2\pi)^2} 
	  \int_0^\infty b_e\rd b_e \ K_1(m_eb_e) \int_0^\infty b_\pi \rd b_\pi \ I(b_\pi,  Q_\|)
	  \nonumber \\ &&
	 \times  2\pi \int_0^{2\pi} \rd \phi 
	  \, J_0\left(\sqrt{ X} \right)\{-\sin\phi\}\exp[i\chi_C({B}_e]  , \label{Lampffint}  
\end{eqnarray} 
with $X$ and $B_e$ defined by
\begin{eqnarray}
	X&=&  (q_\bot b_e)^2+(Q_\bot b_\pi)^2
	             -2q_\bot b_eQ_\bot b_\pi \cos\phi  ,\\
	B_e^2&=&    b_e^2+b_\pi^2+2b_eb_\pi \cos(\phi+\phi_q-\phi_Q). \label{Phase-chi}      
\end{eqnarray}
For general charge and matter distributions $I(b_\pi,  Q_\|)$ is  defined by 
the integral
\begin{equation}
	I(b_\pi,  Q_\|)=  \frac{-2\pi i}{Q_\|} \int_{-\infty}^{\infty} \rd z 
	\, \left[	\hat{\mathbf{b}}_\pi\cdot\mathbf{E}_A(\mathbf{b}_\pi,z)\right]
	\exp[-iz Q_\|-\half \sigma_\pi' T({b}_\pi,z)].
\end{equation}
For a point-charge distribution and neglect of meson rescattering 
it simplifies to
 \begin{equation}
	I_{pc}(b_\pi,  Q_\|)= K_1(Q_\|b_\pi),
\end{equation}
where for negative values of the argument $K_1(-x)=-K_1(x)$.

In Born approximation, i.e.~for a nucleus point-charge distribution and with neglect 
 of electron and pion distortion, 
the amplitudes $K(\mathbf{q},\mathbf{Q})$ 
and $L(\mathbf{q},\mathbf{Q})$ reduce to 
\begin{eqnarray}
	K_{B}(\mathbf{q},\mathbf{Q})&=&  
	  \frac{{q}_\bot}{\mathbf{q}_\perp^2+m_e^2}\cdot 
	  \frac{{Q}_\bot}{\mathbf{Q}_\bot^2+Q_\|^2},
	   \label{KampBorn}\\
	   	L_B(\mathbf{q},\mathbf{Q} )&=&0, \label{LampBorn}
\end{eqnarray}
with $-Q_\|=m_\pi^2/2K_\|$. In this case the integrals factorize. In fact, the functional 
dependence on the angles $\phi_q$ and $\phi_Q$ factorizes except in
the Coulomb-phase-shift function.
\newpage
\section{Hadronic production}

The hadronic contribution represents hadronic interactions between
the high-energy photon and the nucleons in the nucleus. Those interactions
can be described as exchange interactions where the exchanged particle, as 
in Fig.~2, is a 
rho or an omega meson. For omega exchange proton and neutron contributions 
add whereas for rho exchange they subtract. In addition, the rho interaction
is weaker and we shall therefore neglect its contribution.
 The structure of the omega-exchange contribution is the same as that 
 of photon exchange, except for
the replacement of the Coulomb potential by the Yukawa potential.
Hence, the nuclear-electric field $\mathbf{E}_A(\mathbf{r})$ is replaced by the 
nuclear-omega field
\begin{equation}
	\mathbf{E}_\omega(\mathbf{r})=\mathbf{\nabla}_b
	 \frac{1}{4\pi i} \int \rd^3 r' \hat{\rho}_{hd}(\mathbf{r}')
	\frac{e^{-m_\omega |\mathbf{r}-\mathbf{r}'|}}{|\mathbf{r}-\mathbf{r}'|}
	\label{E-omeg-def}
\end{equation}
where $A\hat{\rho}_{hd}(\mathbf{r})$ is the hadronic density distribution as seen
by the omega meson. The omega propagator can be displayed by rewriting
this expression as
\begin{equation}
	\mathbf{E}_\omega(\mathbf{r})= \frac{1}{i(2\pi)^3}\, \mathbf{\nabla}_b \int \rd^3 q
	  e^{-i\mathbf{q}\cdot\mathbf{r}} \frac{1}{\mathbf{q}^2+m_\omega^2} S_0(\mathbf{q})
	  	\label{E-omeg-def-imp}
\end{equation}
with $S_0(\mathbf{q})$ the nuclear form factor,
\begin{equation}
	S_0(\mathbf{q})=\int \rd^3 r
	  e^{i\mathbf{q}\cdot\mathbf{r}}\hat{\rho}_{hd}(\mathbf{r}).\label{For-fac}
\end{equation}

Unfortunately, high-energy photoproduction of pseudoscalar mesons by nucleons
is not well described by the meson exchanges. But Guidal et al.~\cite{Laget1} 
 have shown that by Reggeizing those exchanges a good
 description of data can be obtained. Reggeization means replacing the 
 the omega-pole factor 
\begin{equation}
	{\cal P}_\omega= \frac{1}{t-m_{\omega}^2}
\end{equation}
by its Reggeized version, which is
\begin{equation}
	{\cal P}_\omega(s,t)= \left(\frac{s}{s_0}\right)^{\alpha_\omega(t)-1}
	\frac{\pi\alpha_\omega'}{\sin(\pi\alpha_\omega(t))}\cdot\frac{1}{\Gamma(\alpha_\omega(t))}
	\cdot\frac{S_\omega +e^{-i\pi\alpha_\omega(t)}}{2}.
\end{equation}
The signature $S_\omega=1$, and the parametrization 
of the omega trajectory, $\alpha_\omega(t)$, has been determined through comparison 
with photoproduction  data \cite{Laget1}.
 
For an ambituous calculation that wants to consider the full Regge structure 
the definition of the nuclear-omega field  becomes
\begin{equation}
	\mathbf{E}_\omega(\mathbf{r})= \frac{1}{i(2\pi)^3}\, \mathbf{\nabla}_b \int \rd^3 q
	  e^{-i\mathbf{q}\cdot\mathbf{r}}\frac{1}{m_\omega^2} S_0(\mathbf{q})
	   {\cal P}_\omega(\mathbf{q}^2)/{\cal P}_\omega(\mathbf{q}^2=0)\label{Eomega}
\end{equation}
with $S_0(\mathbf{q})$ the nuclear form factor of Eq.(\ref{For-fac}). 
There could also be other form factors besides the Regge factor. 
The $\mathbf{q}^2$-dependence of the  Regge factor is much weaker than that of
the nuclear factor. Neglecting it altogether leads to a simple expresion for the
nuclear-omega field, 
\begin{equation}
		\mathbf{E}_\omega(\mathbf{r})=\frac{1}{i 
		m_\omega^2} \, \mathbf{\nabla}_b
		 \hat{\rho}_{hd}(\mathbf{r}).
	\label{E-omeg-simple}
\end{equation}

The omega-exchange amplitude is decomposed as follows;
\begin{eqnarray}
	{\cal M}_{\omega}&=&- i {\cal N}_{\omega}\,
	 \bigg[ 	G_{\omega}(\mathbf{q},\mathbf{Q}) 
	   -   H_{\omega}(\mathbf{q},\mathbf{Q})\ 
	   i\bm{\sigma}\cdot \hat{k}_1 \bigg],
	   	 \label{Mom-decomp} \\
	{\cal N}_{\omega}&=&A\,\frac{\alpha g_{\omega\pi\gamma}g_{\omega NN}}{4\pi  m_\pi}
	  \, m_\omega^2{\cal P}_\omega(0).
	\label{Nom-factor}
\end{eqnarray}
Pure omega exchange corresponds to $ m_\omega^2{\cal P}_\omega(0)=-1$.
The functions of Eq.(\ref{Mom-decomp}) are defined in complete analogy with
the two-photon exchange amplitudes, 
\begin{eqnarray}
	G_\omega(\mathbf{q},\mathbf{Q})&=&
	  \int\rd^3r_e \int\rd^3r_\pi\,  e^{i\mathbf{q}\cdot(\mathbf{r}_e -\mathbf{r}_\pi)}
	    e^{-i\mathbf{Q}\cdot\mathbf{r}_\pi}  
	    \exp[i\chi_C({b}_e)-\half \sigma_\pi' T(\mathbf{b_\pi},z_\pi)]\nonumber \\	 &&   
	   \times  \left[\mathbf{E}_e(\mathbf{r}_e -\mathbf{r}_\pi)
	      \times \mathbf{E}_\omega(\mathbf{r}_\pi)\right]\cdot \hat{k}_1  , 
	      \label{Gom-general}\\
	   H_\omega(\mathbf{q},\mathbf{Q})&=&
	  \int\rd^3r_e \int\rd^3r_\pi\,  e^{i\mathbf{q}\cdot(\mathbf{r}_e -\mathbf{r}_\pi)}
	    e^{-i\mathbf{Q}\cdot\mathbf{r}_\pi}  
	    \exp[i\chi_C({b}_e)-\half \sigma_\pi' T(\mathbf{b_\pi},z_\pi)]\nonumber \\	 &&   
	     \times [\mathbf{E}_e(\mathbf{r}_e -\mathbf{r}_\pi)
	      \cdot \mathbf{E}_\omega(\mathbf{r}_\pi)] .\label{Hom-general}   
\end{eqnarray} 

The definitions of $G_\omega(\mathbf{q},\mathbf{Q})$ and $H_\omega(\mathbf{q},\mathbf{Q})$ 
in Eqs (\ref{Gom-general}) and (\ref{Hom-general}) parallel the definitions 
of $G(\mathbf{q},\mathbf{Q})$ and $H(\mathbf{q},\mathbf{Q})$ in 
Eqs (\ref{G-general}) and (\ref{H-general}). To get from the latter to the former 
we replace $\mathbf{E}_A(\mathbf{r}_\pi)$ by $\mathbf{E}_\omega(\mathbf{r}_\pi)$. 
In exactly the same way we define $K_\omega(\mathbf{q},\mathbf{Q})$ and
$L_\omega(\mathbf{q},\mathbf{Q})$ to get the decomposition
\begin{eqnarray}
G_\omega(\mathbf{q},\mathbf{Q}) &=& K_\omega(\mathbf{q},\mathbf{Q})\ 
	 (\hat{\mathbf{q}}_\bot\times\hat{\mathbf{Q}}_\bot)\cdot\hat{k}_1  
	 + L_\omega(\mathbf{q},\mathbf{Q})\ \hat{\mathbf{q}}_\bot\cdot\hat{ \mathbf{Q}}_\bot 
	       ,\label{Gampomegdecomp}	\\
H_\omega(\mathbf{q},\mathbf{Q} )&=& K_\omega(\mathbf{q},\mathbf{Q})\ 
      \hat{\mathbf{q}}_\bot\cdot\hat{ \mathbf{Q}}_\bot - 
	  L_\omega(\mathbf{q},\mathbf{Q}) \ (\hat{\mathbf{q}}_\bot\times\hat{\mathbf{Q}}_\bot)\cdot\hat{k}_1 
	    \label{Hampomegdecomp}. 	
\end{eqnarray}
However, the replacement of $\mathbf{E}_A(\mathbf{r}_\pi)$ by $\mathbf{E}_\omega(\mathbf{r}_\pi)$
makes the structure functions connected with omega exchange quite different from those
connected with photon exchange. In general, $L_\omega(\mathbf{q},\mathbf{Q})$ is small and
may be neglected. For more details see the Appendix.
%
%
%
\newpage
\section{Factorization}\label{five}
%

The formulae given so far apply to arbitrary nuclei and 
arbitrary momentum transfers $\mathbf{Q}_\bot$ and $\mathbf{q}_\bot$,
as long as they remain much smaller than the longitudinal momenta 
$k_1$ and $K$. The tricky point in their evaluation is that the 
integrations over $\mathbf{r}_e $ and $\mathbf{r}_\pi$
are intertwined, as the electron interacts  with both
nucleus and meson. However, in some circumstancies the integrations 
factorize.

Consider production of eta mesons at 11 GeV, and start with
the Coulomb terms $K(\mathbf{q},\mathbf{Q})$ and $L(\mathbf{q},\mathbf{Q})$
 of Eqs (\ref{Kampffint}) and (\ref{Lampffint}). 
The cut-off in the $b_e$ integration is set by the inverse of the electron mass 
$1/m_e\approx 390$ fm, and the cut-off in the $b_\eta$  integration by
the inverse of of the longitudinal momentum transfer 
$2K_\|/m_\eta^2\approx 14$ fm. The only coupling between the
$b_e$ and $b_\eta$ dependencies is in the argument, Eq.(\ref{Phase-chi}), of the 
Coulomb-phase function.
In view of the small ovelap region,  we may here neglect the dependence on $b_\eta$.
As a result, the integrands of Eqs (\ref{Kampffint}) and
 (\ref{Lampffint})  factorize, 
and the function $L(\mathbf{q},\mathbf{Q})$ vanishes after integration over the $\phi$ variable.
 These arguments 
are weakened when we consider momentum transfers 
$\mathbf{q}_\bot$ so large that there are strong oscillations in
the $b_e$ integrand. Then the main contributions to the Coulomb integral
come from regions much closer to the nucleus.

In the factorized approximation
\begin{eqnarray}
     L(\mathbf{q},\mathbf{Q})&=& 0\\
	   K(\mathbf{q},\mathbf{Q})&=& F_C(\mathbf{q})F_A(\mathbf{Q}), 
	   \label{K-subdiv}   
\end{eqnarray} 
with the functions $F_C(\mathbf{q})$ and 
$F_A(\mathbf{Q})$  defined by 
\begin{eqnarray}
	F_C(\mathbf{q})&=&
	\int\rd^3r_e \,  e^{i[\mathbf{q}\cdot\mathbf{r}_e+\chi_C({b}_e)]}\,
	    \hat{\mathbf{q}}_\bot\cdot\mathbf{E}_e(\mathbf{r}_e)
	   ,     \label{FC-def}\\
	   F_A(\mathbf{Q})&=&
     \int\rd^3r_\eta\,  e^{-i\mathbf{Q}\cdot\mathbf{r}_\eta}
	    \hat{\mathbf{Q}}_\bot\cdot\mathbf{E}_A(\mathbf{r}_\eta)
	     \exp[-\half \sigma_\eta' T(\mathbf{b_\eta},z_\eta)].
	  \label{N-def} 
\end{eqnarray} 
The function $F_C(\mathbf{q})$ describes  Coulomb scattering of the electron, and for a 
 nuclear-point-charge distribution it can be integrated exactly, Ref.\cite{GFI,FT}.
The $F_A(\mathbf{Q})$ is the Coulomb contribution (one-photon exchange) to 
 eta-nucleus photoproduction.

In the hadronic contribution the integration over $b_\eta$ is limited
to the nuclear region. Thus, in this case the integrals always factorize,
so that $L_\omega(\mathbf{q},\mathbf{Q})=0$ and 
\begin{eqnarray}
	   K_\omega(\mathbf{q},\mathbf{Q})&=& F_C(\mathbf{q})F_\omega(\mathbf{Q}), 
	   \label{K-om-fac}   \\
	   F_\omega(\mathbf{Q})&=&
     \int\rd^3r_\eta\,  e^{-i\mathbf{Q}\cdot\mathbf{r}_\pi}
	    \hat{\mathbf{Q}}_\bot\cdot\mathbf{E}_\omega(\mathbf{r}_\eta)
	     \exp[-\half \sigma_\eta' T(\mathbf{b_\eta},z_\eta)].
\end{eqnarray} 

The sum of Coulomb and hadronic contributions can be written as
\begin{equation}
	G(\mathbf{q},\mathbf{Q})+G_\omega(\mathbf{q},\mathbf{Q})=
	\sum_\lambda \bigg[\hat{\mathbf{q}}_\bot\cdot \mathbf{e}_\lambda\, F_C(\mathbf{q})\bigg]
	\bigg[ (\hat{\mathbf{Q}}_\bot\times\hat{k}_1)\cdot\mathbf{e}_\lambda\,
	\bigg\{ F_A(\mathbf{Q})+F_\omega(\mathbf{Q})\bigg\}\bigg] ,
	\label{EtaG-decomp}
\end{equation}
where the sum runs over the two photon polarization vectors orthogonal
to $\hat{k}_1$, or $\hat{q}$. 
For the sum $H(\mathbf{q},\mathbf{Q})+H_\omega(\mathbf{q},\mathbf{Q})$ the expression
is the same but with $\hat{\mathbf{q}}_\bot\cdot \mathbf{e}_\lambda$
replaced by $(\hat{\mathbf{q}}_\bot\times\hat{k}_1)\cdot \mathbf{e}_\lambda$.

The two terms in the last bracket of Eq.(\ref{EtaG-decomp}) represent 
the eta-nucleus photoproduction amplitude initiated by photons of polarization $\mathbf{e}_\lambda$. 
The first 
term $F_A(\mathbf{Q})$ is the Coulomb-photoproduction amplitude, the second term $F_\omega(\mathbf{Q})$
the hadronic-photoproduction amplitude.
It is also important to note that the photons radiated by the electron 
are transverse photons, not Coulomb photons.

In pion-electroproduction the hadronic amplitudes  factorize, but the Coulomb 
amplitudes do not. Hence, the pion-electroproduction amplitude will
not have a decomposition as in Eq.(\ref{EtaG-decomp}).
%
%
%
\newpage
\section{Shadowing}\label{six}
%

The hadronic contribution is modified by shadowing, a multple scattering
contribution 
where the initial photon is first converted into a rho meson, which in a 
subsequent collision with another nucleon creates the pion. This
phenomenon is described in detail in Ref.\cite{Yennie}.

In the hadronic term the high-energy photon produces the final-state pion through
omega-meson exchange with a single nucleon. The correspondin amplitude is proportional to
$f_\omega(\gamma N\rightarrow \pi N)$. In the shadowing term 
the high-energy photon first creates a rho meson by diffractive production
on a nucleon. This step is proportional to $f_P(\gamma N\rightarrow \rho N)$.
In the second step the rho meson collides with another nucleon creating
a pi meson through omega-meson exchange. This step is proportional
to the amplitude $f_\omega(\rho N\rightarrow \pi N)$. Now, if the 
hadronic interaction of photons proceeds via the rho meson, 
we expect the relation
\begin{equation}
	f_P(\gamma N\rightarrow \rho N)f_\omega(\rho N\rightarrow \pi N)=
	  f_P(\rho N\rightarrow \rho N)f_\omega(\gamma N\rightarrow \pi N),
	  \label{amprel}
\end{equation}
where $f_P(\rho N\rightarrow \rho N)$ is the diffractive rho-nucleon-scattering
amplitude. We remark that the rho meson is off its mass shell in two
of the amplitudes, one on each side. It is assumed that the off-shell factors
cancel out.

The amplitude relation (\ref{amprel}) leads to a replacement of the omega field, Eq.(\ref{E-omeg-simple}), by
\begin{equation}
	\mathbf{E}_\omega(\mathbf{r})=\frac{1}{i 
		m_\omega^2} \, \mathbf{\nabla}_b \hat{\rho}_{hd}(\mathbf{r})\Bigg\{
		1-\half \sigma'_\rho \int_{-\infty}^z \rd z'n(\mathbf{b},z')e^{i\Delta_\rho(z'-z)}
		 \exp[-\half \sigma'_\rho \int_{z'}^z\rd z''n(\mathbf{b},z'')]\Bigg\},
	\label{E-omeg-shadow}	
\end{equation}
with $n(\mathbf{r})=A\hat{\rho}_{hd}(\mathbf{r})$. The second term inside the brackets
is the shadowing term. The intermediate rho meson is produced at $z'$ and 
the final-state pion at $z$. Between these two points, the distortion of the wave is due
rho-meson scattering.
The longitudinal momentum transfer in the $\gamma N\rightarrow \rho N$ reaction at $z'$ is
$\Delta_\rho=m_\rho^2/2K_\|$. 
The longitudinal momentum transfer in the $\rho N\rightarrow \pi N$ reaction at $z$ is
$\Delta_\pi=(m_\pi^2-m_\rho^2)/2K_\|=-\Delta_\rho-Q_\|$. In the direct hadronic term the 
longitudinal momentum transfer to the pion is $-Q_\|=m_\pi^2/2K_\|$. This phase factor is outside
the omega field $\mathbf{E}_\omega(\mathbf{r})$. So is the pion distortion,  
Eqs (\ref{Gampomegdecomp}) and (\ref{Hampomegdecomp}).
%
%
%

\newpage
\section{Cross-section distributions}
%

The unpolarized-cross-section distribution as derived in Ref.\cite{GFI} reads
\begin{equation}
\frac{\rd \sigma}{\rd^2k_{2\bot}  \rd^2K_{\bot} \rd k_{2\|}}
  = \frac{1}{\pi  K_\|  }\left( \frac{Z \alpha^2g_{\pi\gamma\gamma}}{m_\pi}\right)^2
    \bigg[\, \bigg|  K(\mathbf{q},\mathbf{Q}) + 
       {\cal R} K_\omega(\mathbf{q},\mathbf{Q})\bigg|^2  
       +\bigg|  L(\mathbf{q},\mathbf{Q}) + {\cal R} L_\omega(\mathbf{q},\mathbf{Q})\bigg|^2 \, \bigg],
 \label{Cross-sect-distr-3}
\end{equation}
with ${\cal R}$ the ratio of coupling constants,
\begin{equation}
	{\cal R} = -{\cal N}_{\omega}/{\cal N}_{2\gamma}. \label{coupl-ratio}
\end{equation}
The parameter ${\cal R}$ 
depends only weakly on atomic number so we do not gain relative strength for the Coulomb
term by going to heavier nuclei. The advantage is instead
 that 
the cross-section values themselves grow as $Z^2$.
An interesting feature of Eq.(\ref{Cross-sect-distr-3}) is that in the unpolarized
cross-section distributions
 $K$- and $L$-amplitudes do not  interfere.  The sign of  ${\cal R}$
 must be determined by experiment. 

The transverse momenta  in Eq.(\ref{Cross-sect-distr-3}) 
are restricted to the regions ${q}_\bot,{Q}_\bot \ll k_1,k_2,K$. 
For cross-section distributions  such that 
 ${Q}_\bot R_u\approx 1$,  nuclear structure becomes important and numerical evaluation 
necessary. We have also stressed that depending on the value 
of ${q}_\bot /m_e$ integrals may or may not factorize.
The complicated functional dependences make a general overview  difficult.
For this reason  we concentrate on production of pions and etas in the Coulomb region, 
which is of special importance to the PrimEx experiment. 

\noindent 
{\itshape Pi-meson production.}

Pion electroproduction at 11 GeV/$c$ was investigated in Ref.\cite{GFI},
with emphasis on the double-Coulomb region, where
  $Q_\bot\approx |Q_\| |=0.85$ MeV/$c$ and $q_\bot\approx m_e=0.52$ MeV/$c$.
In this particular case there is, in the Coulomb amplitude, a strong overlap between the impact-parameter domains of pion and electron.  As a consequence 
the electroproduction amplitude does not factorize and the predictions differ
considerably from the Born approximation. Non-factorization means that 
the pion-nucleus-{\itshape photo}production cross section is not a factor of the 
pion-nucleus-{\itshape electro}production cross section.

However, non-factorization does not mean we cannot determine the pion-decay 
constant $g_{\pi\gamma\gamma}$.  We can, as long as we have a reliable theory
for the cross-section distribution
and as long as the hadronic contribution is much smaller than the Coulomb
contribution. 

In Ref.\cite{GFI} the relative size of Coulomb and hadronic amplitudes was estimated.
Here, we assume pure omega exchange and neglect electron and pion distortions.
Then, at the double peak, the ratio of Coulomb to hadronic amplitude strengths becomes
\begin{equation}
	R_\pi=\left[\frac{Z\alpha^2g_{\pi\gamma\gamma}}{m_\pi}\frac{1}{2Q_\|^2}\right]
	\Bigg/\left[\frac{A\alpha g_{\omega\pi\gamma }g_{\omega NN}}{4\pi m_\pi}\frac{1}{m_\omega^2}\right]
	=110. \label{pionratio}
\end{equation}
The numerical value refers to lead nuclei.
Taking Regge exchange instead of omega exchange implies dividing by 
$\mid {\cal P}_\omega(0) m_\omega^2\mid=0.123$ giving a new value, 
$R_\pi=910$. With such a large ratio  it should be easy to isolate the 
Coulomb contribution from the hadronic background,  making a determination of $g_{\pi\gamma\gamma}$ realistic.

Photoproduction of pseudoscalar mesons by protons, including the Coulomb term, was 
investigated in Ref.\cite{Lagetprim}, and photoproduction
by nuclei most recently in Refs \cite{Gevor,PION}. 

\noindent 
{\itshape Eta-meson production.}

Production of eta mesons is at 11 GeV/$c$ simpler to calculate than production of pi mesons.
The double-Coulomb region is now the region where $Q_\bot\approx |Q_\| |=8.5$ MeV/$c$, and
$q_\bot\approx m_e=0.52$ MeV/$c$. Consequently,  overlap between the integration domains 
of the eta- and electron-impact-parameter variables  is small. 
The factorization discussed in Sect.~\ref{five} 
applies and the  cross-section distribution contains the factor
\begin{equation}
	\big|  F_C(\mathbf{q}) \big|^2  
       \big|  F_A(\mathbf{Q}) + F_\omega(\mathbf{Q})\big|^2 .\label{eta-form}
\end{equation}

The first factor in Eq.(\ref{eta-form}) is the squared Coulomb amplitude of the electron. 
Its analytic form is
given in Refs \cite{FT,GFI}. It is not the usual Coulomb-scattering amplitude since it 
describes exchange of transverse photons, not Coulomb photons. 
In the Coulomb factor $F_C(\mathbf{q})$ the longitudinal momentum $q_\|$  
enters in the combination $q_\|/\gamma\approx m_e$.
This factor  exhibits a Primakoff-peak structure in the 
variable $q_\bot$ with 
a peak value  at  $q_\bot\approx m_e$.

The second factor in Eq.(\ref{eta-form}) is the squared eta-nucleus-photoproduction
amplitude. This amplitude  has two parts; the Coulomb amplitude $F_A(\mathbf{Q})$ and
the hadronic amplitude $F_\omega(\mathbf{Q})$. Assuming omega exchange and 
neglecting  electron and pion rescattering 
the ratio $R_\eta$ of Coulomb to hadronic amplitudes is algebraically 
the same as the ratio for pions, Eq.(\ref{pionratio}), but with all pi indices replaced by eta indices. For lead nuclei $R_\eta=1.8$ with omega exchange and  $R_\eta=15$ with Regge exchange. 
The Coulomb amplitude dominates but only weakly. For an accurate determination 
of $g_{\eta\gamma\gamma}$ higher energies are needed.

\noindent 
{\itshape Eta-prime-meson production.}

 At 11 GeV/$c$ the double-Coulomb region in eta-prime production is the region where $Q_\bot\approx |Q_\| |=43$ MeV/$c$, and
$q_\bot\approx m_e=0.52$ MeV/$c$. As for eta-meson production 
the factorization discussed in Sect.~\ref{five} 
applies as well as Eq.(\ref{eta-form}). 
The second factor in Eq.(\ref{eta-form}) is now the squared eta-prime-nucleus-photoproduction
amplitude.  Assuming omega exchange and 
neglecting  electron and pion rescattering 
the ratio $R_{\eta'}$ of Coulomb to hadronic amplitudes is algebraically 
the same as the ratio for pions, Eq.(\ref{pionratio}), but with all pi indices replaced by
 eta-prime indices. For lead nuclei $R_{\eta'}=0.30$ with omega exchange and  $R_{\eta'}=2.5$ with Regge exchange. Consequently, a  determination 
of $g_{\eta'\gamma\gamma}$ is not possible at this energy.
\newpage
\section{Appendix}

Define the normalized Regge propagator as $\hat{{\cal P}}_\omega(\mathbf{q}^2)=
{\cal P}_\omega(\mathbf{q}^2)/{\cal P}_\omega(\mathbf{q}^2=0)$.
With this definition we can write the nuclear-omega field of Eq.(\ref{Eomega}) as
\begin{eqnarray}
	\mathbf{E}_\omega(\mathbf{r})&=& \frac{1}{i(2\pi)^3m_\omega^2}\, \mathbf{\nabla}_b \int \rd^3 q
	  e^{-i\mathbf{q}\cdot\mathbf{r}} S_0(\mathbf{q}^2)
	  \hat{{\cal P}}_\omega(\mathbf{q}^2)\nonumber \\
	  &=& \frac{i}{2\pi^2 m_\omega^2}\,\frac{\mathbf{b}}{r}  \int_0^\infty 
	   q^3\rd q j_1(qr) S_0(\mathbf{q}^2)
	  \hat{{\cal P}}_\omega(\mathbf{q}^2)\label{Eomega-prim}
\end{eqnarray}
If we like the coordinate-space description better we introduce
\begin{equation}
	\hat{{\cal P}}_\omega(\mathbf{r})=\frac{1}{(2\pi)^3}
	    \int \rd^3q e^{-i\mathbf{q}\cdot\mathbf{r}}\hat{{\cal P}}_\omega(\mathbf{r}),
\end{equation}
and get 
\begin{equation}
		\mathbf{E}_\omega(\mathbf{r})= \frac{1}{im_\omega^2}\, \mathbf{\nabla}_b \int \rd^3 r'
		 \hat{\rho}_{hd}(\mathbf{r}-\mathbf{r}')\hat{{\cal P}}_\omega(\mathbf{r}').
\end{equation}
A simplification occurs if we put $\hat{{\cal P}}_\omega(\mathbf{q}^2)=1$, which means
neglecting the angular variation of the Regge factor as compared with the nuclear
form factor. In this approximation $\hat{{\cal P}}_\omega(\mathbf{r})=\delta(\mathbf{r})$,
and we recover Eq.(\ref{E-omeg-simple}).

In momentum space the omega field becomes
\begin{equation}
	\mathbf{E}_\omega(\mathbf{Q})= \int \rd^3r e^{-i\mathbf{Q}\cdot\mathbf{r}}
	\mathbf{E}_\omega(\mathbf{r})
	  =\frac{\mathbf{Q}_\perp}{m_\omega^2} S_0(\mathbf{Q})
\end{equation}
for the simplified Regge case, $\hat{{\cal P}}_\omega(\mathbf{r})=\delta(\mathbf{r})$. 

We end with an alternative method for deriving Eqs (\ref{Gampdecomp}, \ref{Hampdecomp}).
Decompose the unit vector $\hat{{ b}}_e$ along the unit vector $\hat{{ q}}$, which here stands for the 
impact-plane component of ${{\bm q}}$, i.e.
\begin{eqnarray}
	\hat{b}_e&=&[\hat{b}_e\cdot\hat{q}]\,\hat{q}
	  +[\hat{b}_e\cdot(\hat{k}\times\hat{q})]\, \hat{k}\times\hat{q}\nonumber\\
	  &=&\cos (\varphi_e)\, \hat{q} + \sin( \varphi_e)\,  \hat{k}\times\hat{q}.
\end{eqnarray}
We perform the same decomposition for $\hat{{ b}}_\pi$ but along $\hat{Q}$ and with angle 
$\varphi_\pi$. This gives
\begin{eqnarray}
	\hat{b}_e\cdot\hat{b}_\pi&=& \cos\varphi \, \hat{q}\cdot\hat{Q}+
	    \sin\varphi \, (\hat{q}\times\hat{Q})\cdot\hat{k}\\
	 (\hat{b}_e\times\hat{b}_\pi)\cdot\hat{k}&=&-   \sin\varphi\, \hat{q}\cdot\hat{Q}
	  +\cos\varphi \, (\hat{q}\times\hat{Q})\cdot\hat{k},
\end{eqnarray}
with $\varphi=\varphi_e-\varphi_\pi$.


\end{document}